\documentclass[11pt,letterpaper]{article}
\pdfoutput=1
\usepackage{jheppub}
\usepackage{bbm}
\usepackage{mathrsfs}
\usepackage{slashed}
\usepackage{caption}
\usepackage{epstopdf}
\usepackage[normalem]{ulem}
\usepackage[bottom]{footmisc}
\usepackage{subcaption}
\usepackage{bbold}
\usepackage{titlesec}
\usepackage{threeparttable}
\usepackage{booktabs}
\usepackage{changepage}
\usepackage[utf8]{inputenc}
\usepackage{dsfont}
\usepackage{comment} 

\usepackage{grffile}
 
\usepackage{graphicx}  
\usepackage{dcolumn}   
\usepackage{bm}        
\usepackage{amssymb}   
\usepackage{setspace}
\usepackage{amsmath, amssymb, setspace}
\usepackage{array}
\usepackage{booktabs}
\usepackage{caption}
\usepackage{indentfirst}
\usepackage{float}
\usepackage{lmodern}
\usepackage{multirow}
\usepackage{soul}
\usepackage[normalem]{ulem}
\usepackage{xcolor}

 \usepackage{cancel}

\newcommand\be{\begin{equation}}
\newcommand\ee{\end{equation}}

\newcommand\bea{\begin{eqnarray}}
\newcommand\eea{\end{eqnarray}}

\title{\huge{Cosmology-Independent Constraints on Irreducible Magnetic Monopole Background}}

\author[a,b,c]{Syuhei Iguro,} 
\author[d]{Varun Mathur,} 
\author[d]{Ian M. Shoemaker,}
\author[e,c,f,g]{Volodymyr Takhistov}

\affiliation[a]{Institute for Advanced Research (IAR), Nagoya University, Nagoya 464--8601, Japan}
\affiliation[b]{Kobayashi-Maskawa Institute (KMI) for the Origin of Particles and the Universe, Nagoya University, Nagoya 464--8602, Japan}
\affiliation[c]{Theory Center, Institute of Particle and Nuclear Studies (IPNS), High Energy Accelerator Research Organization (KEK), Tsukuba 305-0801, Japan}
\affiliation[d]{Center for Neutrino Physics, Department of Physics, Virginia Tech University, Blacksburg, VA 24601, USA}
\affiliation[e]{International Center for Quantum-field Measurement Systems for Studies of the Universe and Particles (QUP, WPI),
High Energy Accelerator Research Organization (KEK), Oho 1-1, Tsukuba, Ibaraki 305-0801, Japan} 
\affiliation[f]{Graduate University for Advanced Studies (SOKENDAI), \\
1-1 Oho, Tsukuba, Ibaraki 305-0801, Japan}
\affiliation[g]{Kavli Institute for the Physics and Mathematics of the Universe (WPI), UTIAS, \\The University of Tokyo, Kashiwa, Chiba 277-8583, Japan}

\abstract{
We present a novel mechanism for the irreducible production of magnetic monopoles from interactions of cosmic rays and interstellar medium (ISM). Resulting monopoles drain energy from galactic magnetic fields, disrupting their formation and sustainability. We generalize conventional Parker bounds
to monopoles with extended energy spectrum and, considering cosmic ray ISM monopole production, set novel constraints from disruption of Milky Way Galactic magnetic fields and their seeds. Further, we set first constraints on disruption of galactic magnetic fields and their seeds of Andromeda galaxy, with results being competitive with distinct existing bounds. Unlike Parker limits of previous works that relied on cosmological monopoles, our constraints are independent of cosmological monopole production or their primordial abundance. Besides, we estimate new constraints on dipole magnetic moments generated from cosmic ray ISM interactions. We discuss implications for monopoles with generalized magnetic charges.
}

\begin{document}
\preprint{KEK-QUP-2024-0026, KEK-TH-2671, KEK-Cosmo-0365}
 \maketitle
\flushbottom

\section{Introduction}

Magnetic monopoles carrying isolated magnetic charges have been hypothesized for decades and their existence would symmetrize Maxwell's equations with respect to electric and magnetic fields. Concrete formulation of point-like monopoles in 1931 by Dirac~\cite{Dirac:1931kp} showed that monopoles would explain why
fundamental electric charge $e$ is observed to be quantized. The Dirac charge quantization condition establishes a simple relation $eg = n/2$ with magnetic charge $g$, considering natural units $c = \hbar = 1$ and integer
$n$. Taking $n=1$, this implies an elementary Dirac magnetic charge of  
$g_{D} \equiv \frac{\hbar c}{2e} \simeq 68.5e$ (in Gaussian units where $\alpha = e^{2}/(\hbar c)$).
The mass and spin of the magnetic monopole
within Dirac’s model are not fixed parameters. 
Subsequently, it was established that composite magnetic monopoles can generically arise in Grand Unified Theories (GUTs) of unification of forces~\cite{tHooft:1974kcl,Polyakov:1974ek}, where monopoles can appear after spontaneous symmetry breaking of larger gauge group and resulting in their masses being associated with such high energy scales (i.e. $\sim 10^{16}$~GeV in case of GUTs). These considerations are particularly relevant for monopole production at higher energies in the early Universe and associated with phase transitions~(e.g.~\cite{Guth:1979bh}).
Despite decades of searches, monopoles remain undetected and continue to be a fundamental target for probing new physics. 
 
Magnetic monopoles have been associated with rich phenomenology (see e.g.~\cite{Mavromatos:2020gwk} for review).  
This includes catalysis of proton decay through Callan-Rubakov effects~\cite{Rubakov:1981rg,Rubakov:1983sy,Callan:1982ac} and related signature searches in large-volume experiments such as Super-Kamiokande~\cite{Super-Kamiokande:2012tld}, emission of Cherenkov radiation in large experiments such as IceCube~\cite{IceCube:2015agw} as well as ionizing deposits by accelerated relativistic cosmogenic magnetic monopoles in experiments like MACRO~\cite{MACRO:2002jdv}. Strong so-called ``Parker'' limits~\cite{Parker:1970xv} on fluxes of magnetic monopoles arise from considerations of their effects on galactic magnetic fields, which are drained of energy as present magnetic monopoles are accelerated and hence are disrupted faster 
than the dynamo mechanisms can regenerate them.
The presence of a magnetic monopole population can thus disrupt Milky Way's $\sim \mu$G Galactic magnetic fields~\cite{Parker:1970xv,Turner:1982ag}. 
Even more sensitive so-called ``extended Parker'' bounds can be placed on astrophysical monopole fluxes by considering disruption of initial early Milky Way's seed magnetic fields that are subsequently thought to be amplified to present values~\cite{Adams:1993fj}. Versions of Parker bounds include considerations of magnetic monopole fluxes in the context of intracluster magnetic fields~\cite{Rephaeli:1982nv} as well as initial seed magnetic fields within protogalaxy collapse~\cite{Lewis:1999zm}, and recently galactic magnetic fields of Andromeda (M31)~\cite{Zhang:2024mze}.
Previous analyses of magnetic monopole astrophysical Parker bounds\footnote{Here, we make a distinction with monopole bounds considered for primordial magnetic fields and early Universe monopole production~\cite{Kobayashi:2022qpl}.} have typically assumed mono-energetic GUT mass-scale magnetic monopoles that are of cosmogenic origin from the early Universe, with lack of generic alternative astrophysical monopole production sources.

However, as recently investigated in a range of works, light monopoles can appear with masses $m \ll 10^{16}$~GeV that are far below unification scales and that can even be around electroweak scales\footnote{Monopoles with masses below unification scales can also survive early Universe inflation, see e.g.~\cite{Chakrabortty:2020otp} for an example in case of intermediate-mass monopoles.}~\cite{Kephart:2017esj,Cho:1996qd,Cho:2013vba,Ellis:2016glu,Arunasalam:2017eyu,Ellis:2017edi,Arai:2018uoy}. Strong bounds on light magnetic monopoles arise from analyses of their direct production at colliders, such as proton-proton $pp$ collision ATLAS~\cite{ATLAS:2019wkg,ATLAS:2023esy,ATLAS:2024nzp} experiment and  MoEDAL~\cite{MoEDAL:2019ort,MoEDAL:2021mpi,MoEDAL:2023ost} experiment  that also considered heavy-ion (Pb-Pb) collisions~\cite{MoEDAL:2021vix,MoEDAL:2024wbc} and historic searches (see e.g.~\cite{Patrizii:2015uea} for review)
including $pp$ collisions at Intersecting Storage Rings (ISR)~\cite{Giacomelli:1975xy,Hoffmann:1978mp}, proton-antiproton collisions
$p \overline{p}$ at CDF~\cite{CDF:2005cvf} and combined CDF with D0~\cite{Kalbfleisch:2003yt},
electron-positron $e^+e^-$ collisions at PEP~\cite{Fryberger:1983fa}, PETRA~\cite{Musset:1983ii}, TRISTAN~\cite{Kinoshita:1989cb}, OPAL~\cite{Pinfold:1993mq}, MODAL~\cite{Kinoshita:1992wd}
as well as $e^+p$ collisions at HERA~\cite{H1:2004zsc}.
Intriguingly, light magnetic monopole production has been recently investigated from cosmic ray collisions with the atmosphere\footnote{Such ``atmospheric fixed target experiment'' has played an instrumental role in probing fundamental physics historically, including the discovery of neutrino oscillations~\cite{Super-Kamiokande:1998kpq}.}~\cite{Iguro:2021xsu}. 
Unlike production in man-made colliders, atmospheric collisions provide isotropic monopole flux with an extended energy spectrum for all terrestrial experiments. This not only allows cross-correlated searches but also sets leading bounds on TeV-scale monopoles~\cite{Iguro:2021xsu}. 

In this work we investigate a novel scenario of magnetic monopole production from interactions of cosmic rays with interstellar medium (ISM). 
Unlike previous studies of magnetic monopoles that considered cosmogenic magnetic monopole flux, the astrophysical flux of magnetic monopoles produced from cosmic ray interstellar medium collisions is entirely independent of their cosmic relic abundance and early Universe production mechanisms. Resulting monopoles can significantly impact galactic magnetic fields, preventing their formation and disrupting them. By analyzing these effects in detail, we set multiple new cosmology-independent Parker-like bounds on light magnetic monopoles. Further, the novel flux of such magnetic monopoles can also contribute directly to signatures in experiments, although we leave detailed analyses of this for future investigations.

The paper is organized as follows. In Sec.~\ref{sec:cross} we describe a production mechanism of magnetic monopoles from cosmic ray interactions with ISM. In Sec.~\ref{sec:parker} we generalize conventional Parker-like bounds to account for monopoles with extended energy spectrum and set new limits from disruption of Galactic magnetic fields considering cosmic ray ISM production. In Sec.~\ref{sec:extended} we analyze how monopole flux from cosmic ray ISM interactions affects an evolution of galactic seed magnetic fields, setting new extended Parker-like bounds. Then in Sec.~\ref{sec:gencharge} we discuss implications for general magnetic charges. In Sec.~\ref{sec:dipole} we estimate novel Parker-like bounds on magnetic dipoles. Finally in Sec.~\ref{sec:conc} we summarize our findings and conclude.  

\section{Monopoles from Cosmic Ray ISM Interactions}
\label{sec:cross}

As cosmic rays ubiquitously traverse galaxies, their interactions with the interstellar medium (ISM) can lead to copious secondary particle emissions. As we discuss, cosmic ray interactions with ISM can act as an efficient novel source of magnetic monopoles. This new astrophysical production mechanism provides a complementary source to the monopole production arising from cosmic ray collisions with Earth's atmosphere, which has been recently explored in Ref.~\cite{Iguro:2021xsu}. Unlike atmospheric collisions that generate terrestrial monopoles, cosmic ray-ISM interactions yield an astrophysical monopole flux that permeates the galaxy. This distributed flux can significantly influence astrophysical phenomena and observables, such as the behavior of galactic magnetic fields, as we will demonstrate below.
Intriguingly, our mechanism also offers a continuous astrophysical source of magnetic monopoles, distinct from the cosmogenic monopole flux and independent of the primordial monopole abundance typically considered in previous studies.  

To compute the resulting astrophysical monopole flux in our scenario, we must examine monopole production from cosmic ray interactions. Based on perturbative considerations, a production of gauge composite monopoles with additional substructure can be expected to be significantly suppressed, similar to the case of man-made colliders, due to their composite nature~\cite{Drukier:1981fq}. In contrast, point-like Dirac magnetic monopoles do not suffer from this suppression in principle.
Following the conventional treatment of monopole production in collider searches (e.g.~\cite{Baines:2018ltl,ATLAS:2024nzp,MoEDAL:2021mpi,OPAL:2007eyf}) we consider Dirac monopole production within the framework of an effective field theory based on electric-magnetic duality. 

Since cosmic rays are primarily composed of energetic protons and ISM predominantly consists of hydrogen in its neutral or ionized forms, we focus our analysis on monopole production in the context of $pp$ collisions. This approximation captures the dominant contribution to monopole production. In this context of electric-magnetic duality, one can consider perturbative cross-sections for monopole production derived by analogy with their electric counterparts. 
However, due to the lack of proper theory and non-perturbative couplings, robust predictions of monopole production cross-sections remain challenging. 
Hence, we will estimate benchmark scenarios for the monopole production cross-sections by considering relevant tree-level processes, particularly Drell-Yan production through virtual photon $\bar{q}q \rightarrow \gamma^{*} \rightarrow M \overline{M}$ and photon fusion $\gamma^{*} \gamma^{*} \rightarrow M \overline{M}$ as described in Ref.~\cite{Baines:2018ltl}. We note that in lepton collider searches analogous processes were also considered, such as $e^+e^- \rightarrow \gamma^{\ast} \rightarrow M \overline{M}$~\cite{OPAL:2007eyf}.
We consider that produced monopoles do not effectively pair-annihilate shortly after production nor form bound states.

\begin{figure}[t]
\centering
\includegraphics[width=0.5\textwidth]{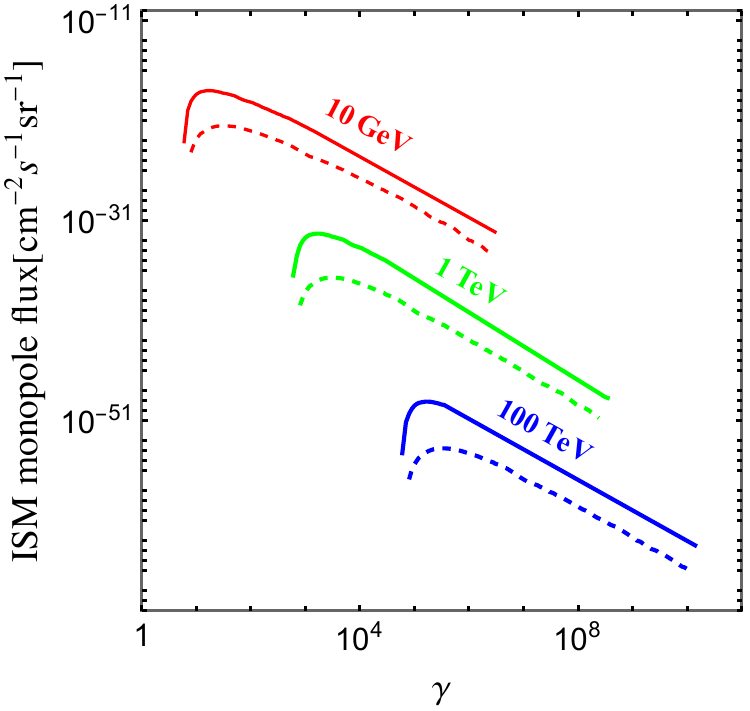 }
    \caption{The differential monopole flux $dF/d\gamma$ produced from cosmic ray ISM collisions for
    10 GeV (red), 1 TeV (green), and 100 TeV (blue) monopole masses as a function of Lorentz factor $\gamma$. Photon fusion (solid) and to Drell-Yan (dashed) production channels are shown.}
    \label{fig:flux}
\end{figure}

To analyze monopole production we carry out detailed Monte Carlo simulations of $pp \rightarrow M \overline{M}$ processes and resulting monopole kinematic distributions, following method of Ref.\,\cite{Iguro:2021xsu}. We employ {\sc\small MadGraph}5 (MG5) version 3.1.0~\cite{Alwall:2014hca} tools with NNPDF31luxQED parton distribution functions~\cite{Bertone:2017bme} and input UFO files corresponding to spin-$1/2$ and velocity independent monopole model~\cite{Baines:2018ltl} that we take as a reference without loss of generality\footnote{The spin-$1/2$ monopole model results in cross-sections around an order larger than spin-$0$ model and an order smaller than spin-$1$ model~\cite{MoEDAL:2019ort}. Velocity-dependent monopole model with coupling $\propto g v$ has also been considered (e.g.~\cite{Epele:2012jn}), which can affect relevant cross-sections by up to a factor of few.
}. This is done for each incident proton energy in the center of mass (COM) frame.
Cosmic rays have extended energy spectra spanning decades of orders of magnitude~(e.g.~\cite{ParticleDataGroup:2024cfk}) and we consider fixed-target collisions with ISM protons. Since monopole pair production requires that the square of the energy in COM is $s \geq 4 m^2$ for monopoles of mass $m$, the resulting lab-frame flux is significantly boosted. 
Here, following method of Ref.~\cite{Iguro:2021xsu}, we treat the overall cross-section normalization as an unknown parameter subject to constraints when compared with data
\begin{equation}
    \sigma(pp \rightarrow M \overline{M})=\kappa  \sigma_{\rm sim},
    \label{eq:sample}
\end{equation}
 where $\sigma_{\rm sim}$ is cross-section resulting from our simulations, and $\kappa$ is the normalization constant. This methodology allows us to consistently compare distinct monopole searches by constraining $\kappa$ for different monopole masses by taking the ratio of the constrained cross-section and simulation predictions at the considered energy.

Cosmic rays permeate ISM, where they induce collisions with gas and dust particles. Cosmic ray sources are generally assumed to be concentrated near the Galactic disk in Milky Way, following a radial distribution similar to that of supernova remnants. At energies $\lesssim10^{17}$~eV, cosmic ray transport in the Galaxy is expected to be well described by a diffusion model, with possible contributions from convection effects~\cite{Strong:2007nh}. We consider 
$d\phi_{\rm CR}/dE_{\rm CR}$ as the differential cosmic ray flux with respect to cosmic ray energy $E_{\rm CR}$. The flux of cosmic rays, $d \phi_{{\rm CR}}/d E_{\rm CR}$
is assumed to be the same as the locally observed spectrum of Galactic cosmic rays at Earth~\cite{ParticleDataGroup:2024cfk}. To a good approximation, the cosmic ray flux spectrum follows $d \phi_{{\rm CR}}/d E_{\rm CR} \propto E_{\rm CR}^{-2.7}$.

With differential cross-section for monopole production $d \sigma_{\rm M}/d\gamma$ for $pp$ collisions with respect to Lorentz factor $\gamma$, the resulting differential astrophysical monopole flux from cosmic ray ISM interactions follows 
\be \label{eq:crismflux}
    \frac{dF}{d\gamma} \simeq \int_{E_{\rm CR}}\frac{d\phi_{{\rm CR}}}{dE_{\rm CR}} \frac{d\sigma_{\rm M}(E_{\rm CR})}{d\gamma}   n^{\perp}_{\rm ISM}.
\ee
Here, $n^{\perp}_{\rm ISM}$ we consider the ISM column density of 
$n^{\perp}_{\rm ISM} \simeq (1~ {\rm cm}^{-3}) \times (1  ~\rm{kpc})$, with characteristic cosmic ray interaction length at kpc-scale. In general, however, the ISM medium and cosmic ray flux in different regions and environments can be expected to induce variations in $n_{\rm ISM}^{\perp}$ within few orders of magnitude across Galaxy. For example, ISM density can vary from $\sim 10^{-26}$ g cm$^{-3}$ in hot medium up to $\sim 10^{-20}-10^{-18}$ g cm$^{-3}$ in dense molecular zones, with an average of around $\sim {\rm few} \times 10^{-24}$ g cm$^{-3}$ ~\cite{Ferriere:2001rg}. Detailed analysis of these effects is beyond our scope and is left for future work.

Here we are interested in the evolution of galactic systems on timescales of magnetic field regeneration of millions to billions of years, as shown in Tab.~\ref{tab:bfield} for Milky Way and Andromeda, corresponding to redshifts $z \lesssim 0.1$. Galactic cosmic rays can originate from distinct sources. Considering characteristic well-studied cosmic ray contributions, from distributions of cosmic star formation rates~\cite{Hopkins:2006bw}, quasars~\cite{jackson:2002} and gamma-ray bursts~\cite{Le:2006pt} we can estimate that over relevant timescales cosmic ray rates are not expected to be altered by more than a factor of few and hence we approximate them as constant in time. 
 
In Fig.~\ref{fig:flux} we display the resulting flux of monopoles produced from cosmic ray ISM interactions for monopoles of different masses, considering Drell-Yan and photon fusion.
Our numerical results indicate that photon fusion always dominates over Drell-Yan production for parameters of interest, in agreement with results of Ref.~\cite{Iguro:2021xsu}. 
 
\section{Galactic Magnetic Fields and Generalized Parker Bound}
\label{sec:parker}

Magnetic monopoles are accelerated in background magnetic fields analogously to electric charges in electric fields. 
Magnetic fields 
permeate a broad variety of astrophysical environments.
Large-scale coherent magnetic $B$-fields in spiral galaxies have been observed to be around $B \simeq \mathcal{O}(\mu$G)~\cite{Widrow:2002ud}.
An astrophysical flux of magnetic monopoles can rapidly drain energy from galactic magnetic fields and disrupt their formation and regeneration. These considerations can result in stringent so-called Parker bounds on general presence of magnetic monopole flux~\cite{Parker:1970xv}, which
have also been extended to initial seed magnetic fields~\cite{Turner:1982ag, Adams:1993fj} and variations considered in distinct scenarios and environments~(e.g.~\cite{Rephaeli:1982nv,Lewis:1999zm,Zhang:2024mze}).

\begin{table}
\begin{center}
\begin{tabular}{c| c| c | c  }
\hline 
\hline
  & Galactic magnetic  & Coherence   & Regeneration     \\ 
    & $B$-field $B_{\rm gal}$ (G) &   length $l_c$ (kpc) &  timescale $\tau$ (yr)  \\ 
  \hline  
 Milky Way & $3 \times 10^{-6}$ & $1 $ & $10^8 $\\  
 \hline
 Andromeda & $5 \times 10^{-6}$ & $10$ &  $10^{10} $ \\
 \hline\hline
 \end{tabular}
 \caption{Estimated galactic magnetic field strength, coherence length and regeneration timescale for Milky Way~\cite{Beck:1995zs} and Andromeda~\cite{Beck:1995zs,Han:1998he,Fletcher:2003ec,Arshakian:2008cx} galaxies.}
 \label{tab:bfield}
\end{center}
\end{table}

The energy gained or lost by magnetic monopoles of charge $g$ in crossing a magnetic field $B$ of coherence length $l_c$~\footnote{Note that actual coherence length  $l_c$ varies and can be $\mathcal{O}(100)$~pc, with effects such as field reversals and turbulence playing a role. We consider $\sim 1$~kpc throughout to obtain approximate bounds.} is
\begin{equation} \label{eq:emag}
    E_{\rm mag}=g B l_c \simeq 3 \times 10^{9} ~{\rm GeV}~\left(\frac{g}{1}\right)~\left(\frac{B}{10^{-6}~{\rm G}}\right)\left(\frac{l_c}{1~{\rm kpc}}\right),
\end{equation}
where in the above we have adopted the characteristic values for the Milky Way Galactic magnetic fields. In Tab.~\ref{tab:bfield} we display estimated galactic magnetic field parameters we consider for the Milky Way~\cite{Beck:1995zs} and Andromeda~\cite{Beck:1995zs,Han:1998he,Fletcher:2003ec,Arshakian:2008cx} galaxies.
For Andromeda, we shall assume that cosmic ray flux does not significantly differ from that of Milky Way. In actuality, a detailed magnetic field structure of galaxies is significantly more complex.

We start by generalizing the conventional Parker bound in two ways, going beyond what was done in Refs.~\cite{Parker:1970xv,Turner:1982ag} to formulate an extended Parker bound. First we will establish the criteria for monopole flux with an extended energy spectrum, breaking
the typical assumption that the entire monopole flux is monoenergetic. Second, we will also generalize our constraints to include relativistic monopoles. 

The Galactic dynamo is considered to be the prevailing mechanism of magnetic field amplification in the Milky Way~\cite{Beck:1995zs}, operating on a timescale of $\tau \simeq 10^{8}$ yr. By requiring that the rate of energy removed by a monopole flux $dF/dE$ does not exceed the Milky Way's magnetic energy budget on the field's regeneration timescale $\tau$ yields   
\begin{equation}
    \frac{1}{\tau} \frac{B^{2}}{8\pi} \frac{4\pi}{3} l_c^{3} > \int ~dE\frac{dF}{dE} 4 \pi l_c^{2} \Delta E, 
    \label{eq:parker1}
\end{equation}
where $\Delta E$ is the average energy gained by a magnetic monopole in the Galaxy. 

\begin{figure}[t]
\centering
\includegraphics[width=0.6\textwidth]{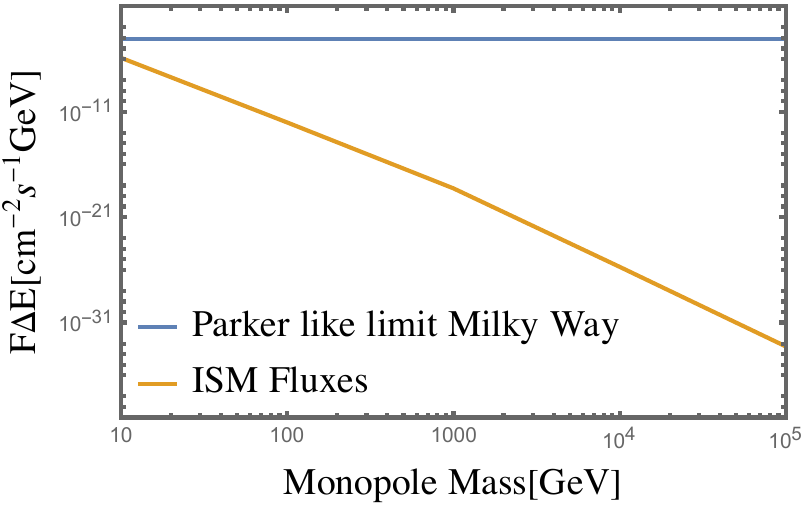}
\caption{
Here we plot the right hand side of Eq.~\eqref{eq:parker} (show in blue) and the left hand side (shown in yellow). To obtain constraints then from this Parker-like bound, we rescale the production cross section until the yellow curve crosses the blue. }
\label{fig:parker}
\end{figure}

As discussed in~Ref.~\cite{Turner:1982ag} for non-relativistic monopoles the energy gained depends on whether the initial monopole energy $E_i$ is small or large compared to $E_{\rm mag}$. In App.~\ref{app}, we extend this analysis to relativistic monopoles. We then find regimes of $\Delta E$  
\be
\Delta E =\begin{cases}
			gBl_{c}~, & E_{i} < E_{\rm mag}\\
            \dfrac{1}{2}  \dfrac{g^{2} B^{2}l_{c}^{2}}{m\gamma \beta^{2}}~, & E_{i} > E_{\rm mag}
		 \end{cases}
\label{eq:monoE}
\ee
where $\beta$ is the monopole velocity and $\gamma$ is the Lorentz factor. 
With this, we can generalize Eq.~\eqref{eq:parker1} to account for both possible contributing energy loss regimes  as
\begin{equation}
    \frac{1}{\tau} \frac{B^{2} l_c}{24\pi} >\left( \int^{E_{\rm mag}} ~dE\frac{dF}{dE} \Delta E_{1} +\int_{E_{\rm mag}} ~dE\frac{dF}{dE} \Delta E_{2} \right) \equiv F \Delta E, 
    \label{eq:parker}
\end{equation}
where we have canceled the constant factor $4 \pi l_c^2$ from both sides,
and where $\Delta E_{1}$ applies in the $E < E_{\rm mag}$ regime, while $\Delta E_{2}$ applies when $E > E_{\rm mag}$.

One may consider if the production of monopole anti-monopole pairs through cosmic ray ISM collisions leads to a net zero energy loss effect on galactic magnetic fields since while monopoles with charge $g$ drain field energy antimonopoles with charge $-g$ will initially decelerate restoring field energy (see e.g. Eq.~\eqref{eq:emag}). However, for small initial monopole energy in the regime of $E_i < E_{\rm mag}$, as described by Eq.~\eqref{eq:monoE}, as anti-monopole decelerates it will eventually come to rest and will instead begin to accelerate draining magnetic field energy that it had temporarily restored. On the other hand, in the regime of $E_i > E_{\rm mag}$ since $\Delta E \propto g^2$ both monopoles and anti-monopoles will also drain magnetic field energy.

As an illustration, let us consider a characteristic example of an astrophysical extended power law flux for magnetic monopoles, $dF/dE= A~ (E/E_{0})^{-n}$ where $A$ is a constant, $E_{0} = 10$ GeV, $n >1$ is positive real and that flux is nonzero for energies $E> E_0$. Let us assume that such a flux exists in the $E<E_{\rm mag}$ regime, such that each monopole gains an energy of $\Delta E = E_{\rm mag} = g B l_c$. Then,
\be 
F_{\rm tot} \equiv \int \frac{dF}{dE} ~dE = \Big(\dfrac{A E_0^n E^{1-n}}{1-n}\Big)\Big|_{E_0}^{\infty} \xrightarrow[n = 2]{}
  E_0~A \lesssim ~10^{-15}~{\rm cm}^{-2}~{\rm s}^{-1} ~{\rm sr}^{-1}~, 
\ee
where after integration we have assumed for illustration the special case $n = 2$ and evaluated the right hand side flux considering Eq.~\eqref{eq:parker1} and magnetic field parameters for Milky Way. Hence, in terms of normalization of the differential monopole flux, this becomes 
\be 
A (n = 2)\lesssim 10^{-16}~{\rm cm}^{-2}~{\rm s}^{-1} ~{\rm sr}^{-1}~{\rm MeV}^{-1}~. 
\ee
 
In Fig.~\ref{fig:parker} we display the left- and right-hand sides of Parker-like bound Eq.~\eqref{eq:parker} as a function of monopole mass
considering computed monopole flux from cosmic ray ISM interactions of Eq.~\eqref{eq:crismflux}. We observe that our scenario excludes monopoles with masses lighter than $M \lesssim 9$ GeV considering Milky Way Galactic magnetic field disruption in Parker-like bounds. Importantly, these limits are independent of monopole cosmic relic abundance. Note that here we are implicitly assuming that Galactic magnetic fields, ISM and cosmic ray interactions are approximately constant on timescales of dynamo field regeneration $\tau \sim 10^{8}$ yr. 

 We briefly comment on the high-monopole mass limit of the Parker-like bound in Eq.~\eqref{eq:parker} that we consider for our scenario.
 In the high-mass limit, the flux distributions originating from cosmic ray ISM interactions as displayed on Fig.~\ref{fig:flux} are dominated by monopole energies that are large compared to characteristic $E_{\rm mag} \simeq 3 \times 10^{9}$~{\rm GeV} of Eq.~\eqref{eq:emag} for Milky Way. Thus, $\Delta E$ appearing in Eq.~\eqref{eq:parker} is expected to be in the $E_i > E_{\rm mag}$ regime described in Eq.~\eqref{eq:monoE}, with $\Delta E = (gBl_{c})^{2}/(2m\gamma \beta^{2})$. Hence,  Eq.~\eqref{eq:parker} becomes
\begin{equation} \label{eq:parkerlim}
    \frac{1}{\tau} \gtrsim   \frac{12\pi g^{2} l_c}{m}    \int \frac{dE}{\gamma \beta^{2}}\left(\frac{dF}{dE}\right) .
\end{equation}

\section{Extended Parker-like Bound on Monopoles}
\label{sec:extended}

Large-scale Galactic magnetic fields observed today are thought to have been seeded by processes in the early Universe, and subsequently amplified via $\alpha-\Omega$ dynamo action in the Galaxy to present strengths~(see e.g.~\cite{annurev:/content/journals/10.1146/annurev-astro-071221-052807}). 
Our analysis of effects of magnetic monopoles from cosmic ray ISM collisions can be extended to earlier times when galactic magnetic fields were being formed from smaller nascent ``seed'' $B_0$ magnetic fields. 
Depending on the origin and evolution details, a variety of seed magnetic fields have been suggested including $B_0 \simeq 10^{-18}$~G as can be expected from Biermann battery mechanism~\cite{Rees:1987}. However, even smaller seed fields such as $B_0 \simeq 10^{-20}$~G have been also suggested (see e.g.~\cite{Enqvist:1998fw} for discussion).
In Ref.~\cite{Adams:1993fj} the original Parker analysis was extended to earlier times wherein the presence of magnetic monopoles could disrupt the seed magnetic fields, which are significantly smaller than Galactic magnetic fields at present time and hence provide additional sensitivity to probing the presence of magnetic monopoles.

While for computing Parker-like bound for our scenario in Eq.~\eqref{eq:parker1} galactic magnetic fields were considered as constant at the present time, we now consider their evolution and dissipation from seed fields over time by generalizing for our scenario the extended Parker-like bound to include monopoles with extended energy spectrum and relativistic effects.
A simplified model of the time-evolution of galactic magnetic fields can be written as
\be 
\frac{dB}{dt} = \gamma_{\rm gal} B - \frac{\gamma_{\rm gal}}{B_{\rm gal}} B^{2} -  12 \pi g~ \int d\gamma~ \left( \frac{dF}{d\gamma} \right)   ~\frac{1}{1+ 2m \beta^{2} \gamma/(gB l_c)},
\label{eq:extendedpaker}
\ee
where $\gamma_{\rm gal} = 1/\tau$ is the rate of magnetic field regeneration due to the galactic dynamo in units of $10^{-8}{\rm yr}^{-1}$,  $t$ is time in units of $10^8$~yr. In Eq.~\eqref{eq:extendedpaker} the second term $\propto B^2$ represents turbulent dissipation, with $B_{\rm gal}$ representing galactic magnetic field (see Tab.~\ref{tab:bfield}), while the last term represents dissipation due to magnetic monopoles. Note that the integral for monopole flux is over the monopole Lorentz boost factor $\gamma$.

\begin{figure*}[t]
    \includegraphics[width=0.5\linewidth]{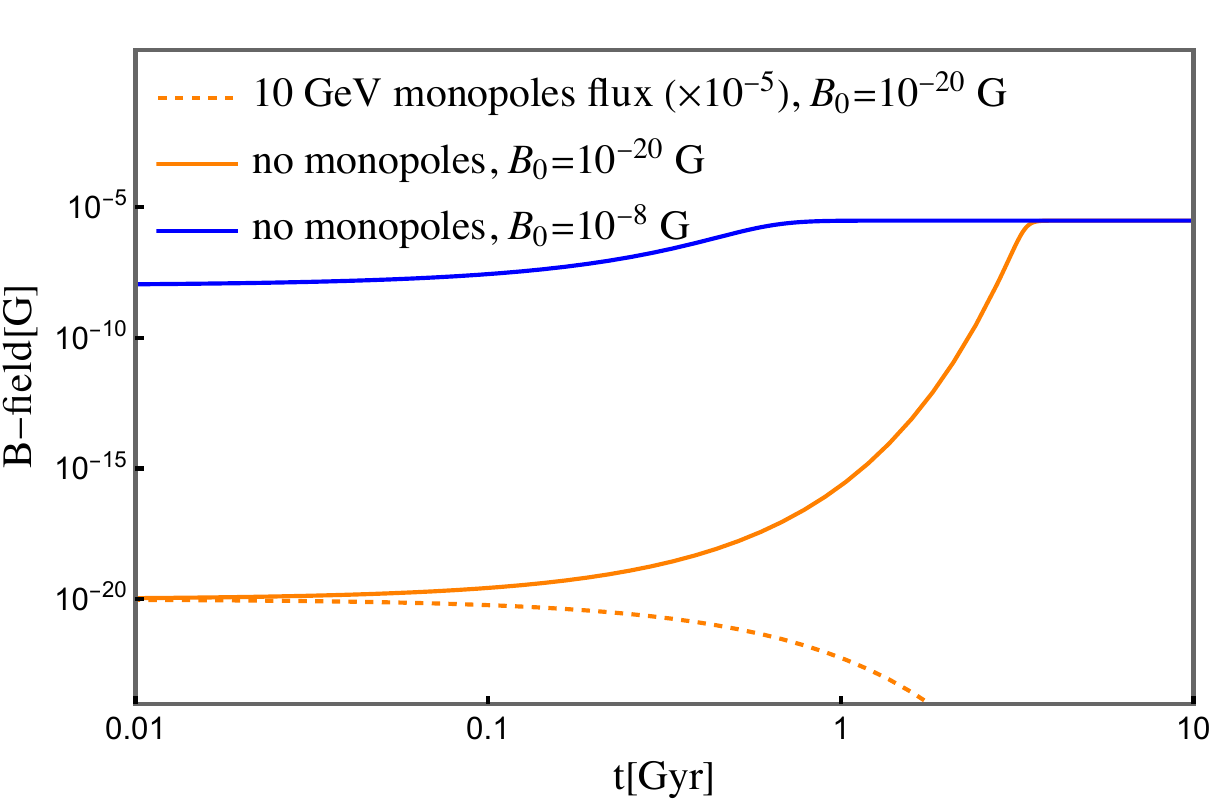}
    \includegraphics[width=.5\linewidth]{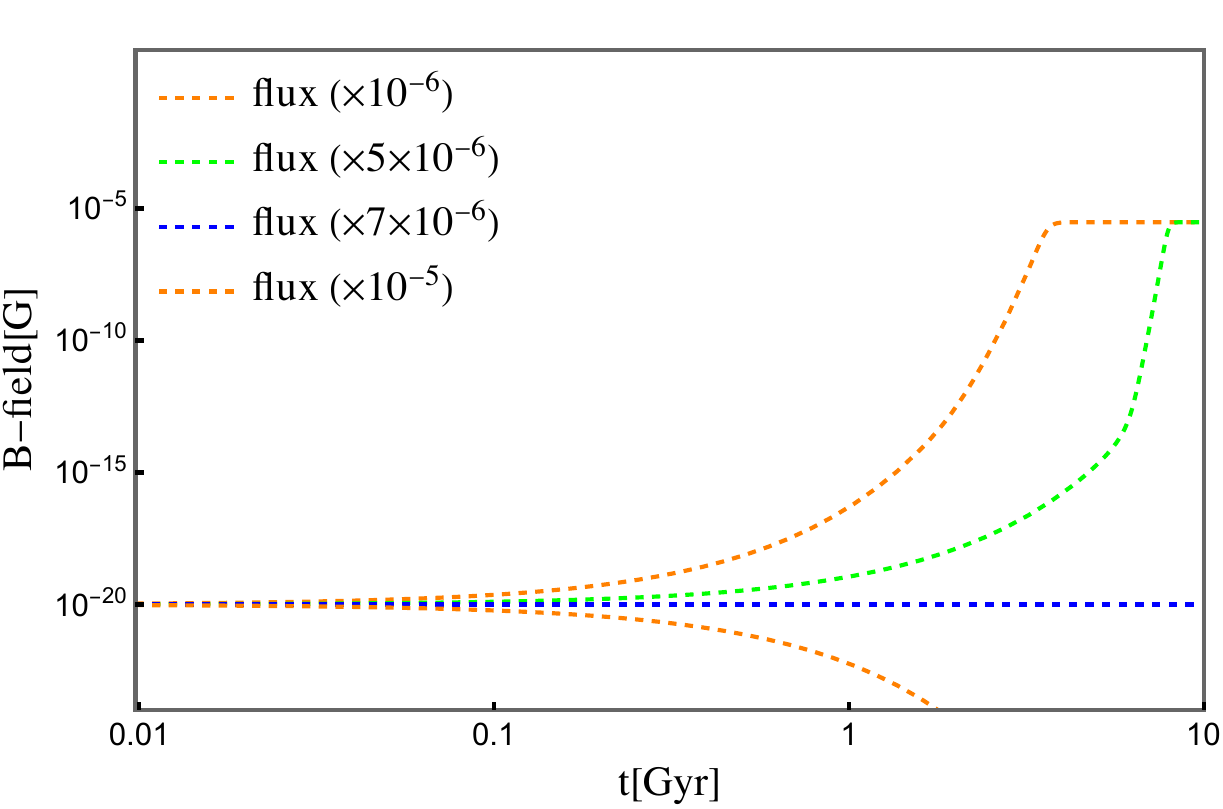}
        \caption{[Left] Magnetic field evolution from solution of Eq.~\eqref{eq:extendedpaker} considering different seed fields $B_0$ with and without effects of magnetic monopoles. [Right] Magnetic field evolution from solution of Eq.~\eqref{eq:extendedpaker} with different fluxes of 10 GeV considering seed field $B_0=10^{-20}$~G. 
        }
\label{fig:extended}
\end{figure*}

\begin{figure}[t!]
\centering
\includegraphics[width=0.85\textwidth]
{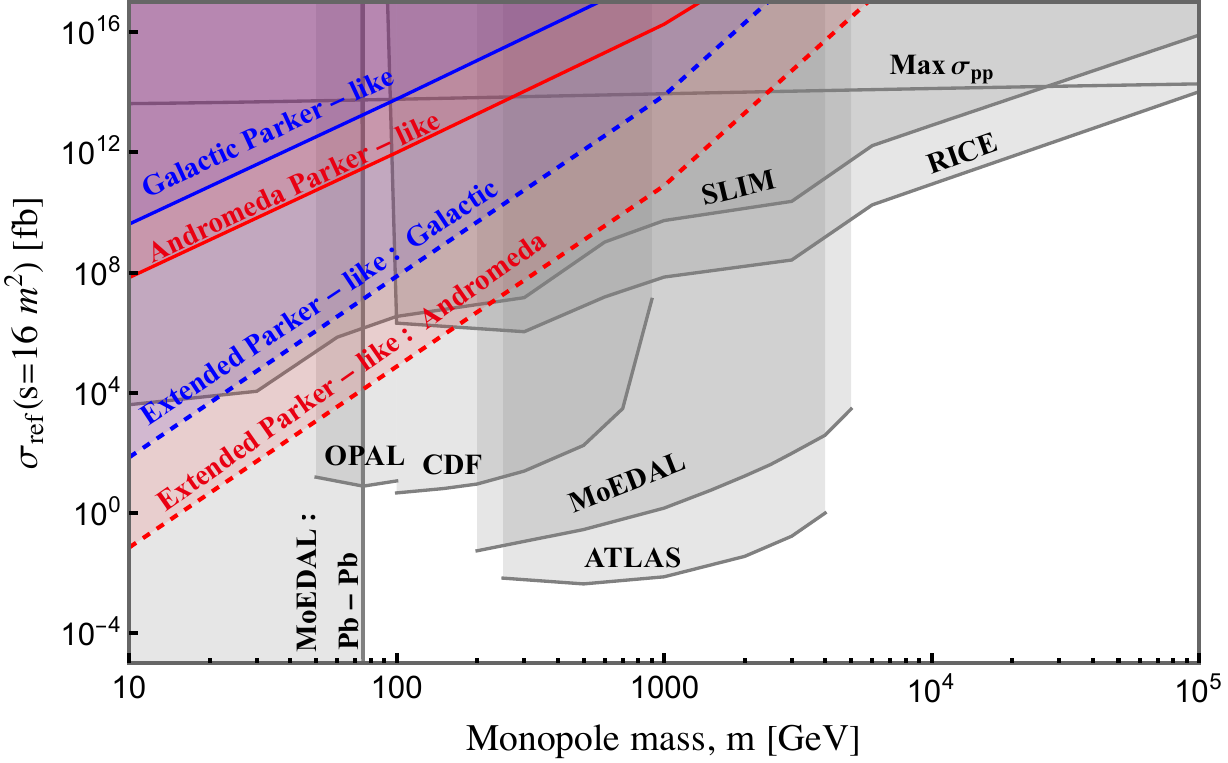}
\caption{
Summary of monopole bounds from the Parker-like (Sec.~\ref{sec:parker}) and extended Parker-like constraints taking the seed field $B_{0} = 10^{-20}$ G (Sec.~\ref{sec:extended}). Existing bounds (gray) are obtained from the combination of constraints: from a search for high ionization energy loss $dE/dx$ in OPAL~\cite{OPAL:2007eyf} and CDF~\cite{CDF:2005cvf}, the MOEDAL~\cite{MoEDAL:2019ort} trapping detector at LHC, ATLAS~\cite{ATLAS:2019wkg}, as well as the RICE~\cite{Hogan:2008sx} and SLIM~\cite{Balestra:2008ps} data reinterpreted in Ref.~\cite{Iguro:2021xsu}. Note that the very strong bounds on monopoles $\lesssim 75$ GeV arise from the Schwinger produced monopoles in Pb-Pb collisions~\cite{MoEDAL:2021vix,MoEDAL:2024wbc}. We also display the maximum total cross section $\sigma_{pp \rightarrow X}$, as parameterized by the
COMPETE collaboration~\cite{COMPETE:2002jcr}. 
} 
\label{fig:bounds}
\end{figure}

We note that a simplified approximate constraint on monopole flux similar to that of Eq.~\eqref{eq:parkerlim} can be obtained from the extended Parker-like bound of Eq.~\eqref{eq:extendedpaker} in the high monopole mass limit.
Considering large mass $m$ and neglecting term $\propto B^2$ for smaller values of $B$, Eq.~\eqref{eq:extendedpaker} can be approximated as
\begin{equation}
   \frac{dB}{dt} \simeq \gamma_{\rm gal} B -12 \pi g \int d\gamma \Big(\frac{dF}{d\gamma}\Big)  \frac{gBl_c}{2m \beta^{2} \gamma} .
\label{eq:approx}
\end{equation}
By requiring that the amplification of magnetic field is always positive and not disrupted by monopoles we can derive a conservative bound on monopole flux. Taking $dB/dt \gtrsim 0$ and definition of $\gamma_{\rm gal}$, we obtain
\begin{equation}
\label{eq:approxbound}
    \dfrac{1}{\tau} \gtrsim \frac{6 \pi g^{2} l_c}{m}\int \frac{d\gamma}{\gamma \beta^{2}}\left(\frac{dF}{d\gamma} \right).
\end{equation}
We observe that the bound of Eq.~\eqref{eq:approxbound} is within a factor of a few of
the Parker-like bound of Eq.~\eqref{eq:parkerlim} for galactic magnetic fields at the present time. Our result explains the agreement with the behavior observed in earlier studies that found Parker and extended Parker bounds to asymptotically approach each other at the high monopole masses (e.g.~\cite{Adams:1993fj,Kobayashi:2023ryr}).

In Fig.~\ref{fig:extended} we illustrate galactic magnetic field evolution by solving Eq.~\eqref{eq:extendedpaker} for monopoles of 10 GeV mass, considering cosmic ray ISM monopole flux as we computed in Fig.~\ref{fig:flux}. We observe that flux of such monopoles can completely disrupt the dynamo amplification of the seed magnetic field depending on their astrophysical flux. Hence, these monopoles are ruled out. We further illustrate in Fig.~\ref{fig:extended} right panel that for lower but still non-negligible flux of magnetic monopoles the seed magnetic fields can be amplified to present day values but in delayed time.

To systematically compare distinct monopole searches, we compare the normalization constant $\kappa$ from Eq.~\eqref{eq:sample} for different monopole masses $m$, as discussed in Ref.~\cite{Iguro:2021xsu}. We consider COM energy $\sqrt{s} =16 m^2$ for each monopole mass and rescale the simulation cross section $\sigma_{\rm sim}$ at that energy by $\kappa$ and compare with our new astrophysical bounds as well as existing laboratory searches.

In Fig.~\ref{fig:bounds} we display novel resulting constraints on the reference cross-section $\sigma_{\rm ref}$ derived from both, our Parker-like analyses considering disruption of present galactic magnetic fields as well as extended Parker-like analyses considering disruption of initial seed galactic magnetic fields $B_0$ in case of Milky Way and Andromeda galaxies, for the scenario of monopole production from cosmic ray ISM interactions. To obtain these limits,
we rescale the computed cosmic ray ISM monopole flux up until it catastrophically disrupts the dynamo and inhibits the generation of $\mathcal{O}(\mu {\rm G})$ galactic magnetic fields observed at present. For extended Parker-like limits, the results depend on the assumed initial seed magnetic field and we illustrate limits considering $B_0 = 10^{-20}$G.

Let us comment that monopoles will generally experience energy losses beyond magnetic field drain as they traverse galactic environment. Considering the monopole stopping power $dE/dx$ due to collisional energy losses for monopoles traveling a distance $x$ in a matter medium is described by modified Bethe-Bloch formula~\cite{PhysRevD.17.229,Patrizii:2015uea} 
\be \label{eq:bethebloch}
\frac{dE}{dx} \simeq \frac{4\pi n_{e} g^{2} e^{2}}{m_{e}} ~\left[\log{\Big(\frac{2m_{e} \beta^{2}\gamma^{2}}{I}\Big)} \right] \simeq 7.4 \times 10^{-3}~{\rm GeV}~{\rm kpc}^{-1}~\left[12+ 2\log{\left(\frac{\beta \gamma}{10^{4}}\right)}\right]~,
\ee
where we used Dirac's quantization condition and $m_e = 0.511$~MeV is the electron mass, with medium electron density $n_e \simeq 1$~cm$^{-3}$ typical of ISM considering neutral hydrogen (HI) regions dominate, $e$ is the electric charge as before and $I \simeq 13.6$ eV is the mean excitation energy of hydrogen. Additional corrections to Eq.~\eqref{eq:bethebloch} such as Bremsstrahlung can further enhance the energy losses~\cite{PhysRevD.17.229,Wick:2000yc}, but are not expected to substantially affect our estimates for relevant parameter space.

\section{General Magnetic Charges}
\label{sec:gencharge}

While our analysis has primarily focused on monopoles with magnetic charges 
$g= g_D$, the new production mechanism and its effects on astrophysical magnetic fields discussed here are applicable to a significantly broader parameter space. This includes monopoles with charges smaller than unity of Dirac charge $g_D$. Recently, there has been growing interest in the phenomenology of monopoles with 
$g \ll g_D$ or fractional magnetic charges~\cite{Brummer:2009cs,Hook:2017vyc,Graesser:2021vkr,Kobayashi:2023ryr}. As we will review below, many of the existing constraints for light monopoles do not apply to such small magnetic charges, further highlighting relevance of our scenario across this extended regime. We note that the Parker-like and extended Parker-like bounds we have derived in this work do not suffer from these caveats, and well-apply with some re-scalings to a broad variety of monopole charges including those below unit Dirac charge.

Strong bounds on light magnetic monopoles with mass $M \lesssim 80$ GeV have been recently set considering Schwinger pair production from magnetic fields arising in Pb-Pb heavy-ion collisions the Large Hadron Collider (LHC) by MoEDAL~\cite{MoEDAL:2021vix,MoEDAL:2024wbc}. The experimental setup utilizes passive detectors composed of aluminum blocks whose large magnetic moment is favorable for binding of magnetic monopoles. After a period of exposure, the passive detector material is then scanned for monopoles with a SQUID magnetometer. The resulting stringent constraints on magnetic monopoles impose that magnetic charge is $g < 45 g_D$. Larger magnetic charges rapidly lose their energy and trapping efficiency decreases when $g_D \gtrsim 6$, although Schwinger production cross-section increases with magnetic charge. A lack of a statistically significant monopole detection robustly excludes monopoles in the detector blocks with $g > 0.5 g_{D}$ at $3 \sigma$, as smaller charged monopoles do not lose sufficient energy and just penetrate the detector. Hence, the constraints from Ref.~\cite{MoEDAL:2021vix,MoEDAL:2024wbc} that we display in Fig.~\ref{fig:bounds} for $M \lesssim 80$ GeV, apply to monopoles with $0.5 g_{D}< g < 45g_{D}$.  

Other stringent bounds besides heavy-ion collisions exist for light magnetic monopoles. In particular, for masses below $\sim 80$ GeV, analysis of Ref.~\cite{Iguro:2021xsu} showed that monopoles produced from cosmic ray atmospheric collisions can be efficiently constrained by historic data from SLIM high altitude nuclear track detector~\cite{Balestra:2008ps} up to $ \sigma_{{\rm ref}} < 10^{4} $ fb at mass of $M=10$ GeV. As displayed in Fig.~\ref{fig:bounds}, our extended Parker-like bounds considering seed galactic magnetic field of $B_{0} = 10^{-20}$ G are comparable to these results. We further note that considering SLIM data the cross-section constraints scale as $\sigma_{{\rm ref}} \propto g^{-4}$, whereas for our Parker-like bounds the scaling follows $\sigma_{{\rm ref}} \propto g^{-3}$ since the generic flux constraint $F \propto g^{-1}$ for monopoles with initial energies small compared to $E_{\rm mag}$ (see e.g. Eq.~\eqref{eq:monoE}). Thus, for $g\ll g_{D}$ the new derived Parker-like bounds can exceed those from the SLIM data for smaller monopole masses.

Other bounds on low-mass magnetic monopoles include $e^+e^-$ collisions at COM energy $\sqrt{s} = 29$~GeV at PEP at SLAC that are valid when $0.3 g_{D}\le g \le 2.9g_{D}$~\cite{Fryberger:1983fa}, $e^+e^-$ collisions at $\sqrt{s} = 34$~GeV at PETRA at DESY which applies to $1.0 g_{D} \le g \le 5g_{D}$~\cite{Musset:1983ii}. The Intersecting Storage Rings (ISR) experiment at CERN used $pp$ collisions to produce a variety of constraints on monopoles\footnote{One of these, which appears as ``ISR1'' in Fig.6 of Ref.~\cite{Patrizii:2015uea}, only appears in proceedings~\cite{Stone:1984vom}.}. The earliest ISR results report limits on low-mass monopoles with $M \le 20$ GeV for magnetic charge $0.4g_{D} < g < 2.5g_{D}$~\cite{Giacomelli:1975xy,Hoffmann:1978mp}. TRISTAN $e^+e^-$ collisions at KEK with $\sqrt{s} = 50-60.8$~GeV constrain $0.2g_{D} \le g \le 2 g_{D}$ for masses $M \le 28.8$ GeV~\cite{Kinoshita:1989cb}. 
Multiple searches for monopoles were also carried out at LEP $e^+e^-$ collider at CERN. Search at OPAL intersection point constrained monopoles with $M \le 45$ GeV and charges $0.9g_{D} \le g \le 3.6 g_{D}$~\cite{Pinfold:1993mq}. HERA at DESY employed $e^{+}p$ collisions for monopole production and SQUID magnetometers to search for trapped monopoles in their beam pipe yielding constraints on monopoles with $M \le 140$ GeV and charges $0.1 g_{D} \le g \le 6 g_{D}$~\cite{H1:2004zsc}.  The MODAL (Monopole Detector at LEP) experiment utilized LEP's $e^{+}e^{-}$ collisions to search for ionization signals from monopoles in plastic track detectors, yielding bounds for $M < 44.9 $ GeV monopoles with charges $0.1 g_{D}\le g  \le 3.6 g_{D}$~\cite{Kinoshita:1992wd}. CDF conducted a search explicitly only for $g= g_{D}$ monopoles excluding monopoles in the mass range $100-900$ GeV~\cite{CDF:2005cvf}. Although these constraints may potentially apply to other magnetic charges, their regime of validity is restricted to be within the 100-900 GeV mass window. Similarly the E882 experiment utilized D0 and CDF data to constrain monopoles with charges $g = (1-6)g_{D}$ up to masses $M \lesssim 200 $ GeV~\cite{Kalbfleisch:2003yt}. 

In summary, our review of variety of bounds outlined in Refs.~\cite{Patrizii:2015uea,Iguro:2021xsu} suggests that none of them definitively apply to monopoles with masses $M \lesssim 80$ GeV and magnetic charges $g \lesssim 0.1 g_{D}$. This is in contrast to astrophysical Parker-like bounds on magnetic monopoles from our scenario of cosmic-ray ISM collisions outlined in this work. We leave a detailed analysis of such monopoles with distinct magnetic charges for future investigation.

\section{Magnetic Dipole Parker-like Bounds}
\label{sec:dipole}

One can extend both the original Parker bound and the Parker-like bounds from cosmic ray ISM collisions to particles with dipole moments\footnote{In Ref.~\cite{Long:2015cza} bounds on neutrino magnetic moments were also considered in the context of primordial magnetic fields.}. This can be achieved by modifying Eq.~\eqref{eq:monoE}, originally derived assuming magnetic monopoles, and extend it appropriately for magnetic dipoles. Considering a uniform galactic magnetic field, energy change will originate primarily from change of dipole orientation and not translational motion, which we can approximate as
\begin{equation}
    \Delta E_{\rm dip} = \mu_{\rm dip}B~,
\end{equation} 
with $\mu_{\rm dip}$ denoting the dipole moment of the new particle.    

Then, we find that revising the Parker limit for magnetic monopoles disrupting galactic magnetic fields results in a bound on magnetic dipoles of
\be \label{eq:fdip}
F_{\rm dip} \lesssim \frac{Bl_c}{24 \pi  \tau \mu_{\rm dip}}~.
\ee
Eq.~\eqref{eq:fdip} can be compared with the Parker bound on magnetic monopoles
\be
F_{\rm mon} \lesssim \frac{B}{24 \pi g \tau }.
\ee
on the dipole moments from LHC searches at low masses $m_{\rm dip} \lesssim 100$~GeV with  $\mu_{\rm dip}\lesssim 10^{-3}~{\rm GeV}^{-1}$~\cite{Fortin:2011hv}. Hence, we find that the resulting Parker-like bound on magnetic dipole flux will be significantly weaker compared to the analogous flux of magnetic monopoles by orders of magnitude
\be
\frac{F_{\rm dip}}{F_{\rm mon}} \simeq 10^{38} ~\left(\frac{g}{1}\right) \left(\frac{ l_{c}}{{\rm kpc}}\right) \left(\frac{10^{-3}~{\rm GeV}^{-1}}{\mu_{{\rm dip}}}\right).
\ee
We note that these are new independent constraints and depending on model may play a relevant role.
We leave detailed further analyses of this for future work.

\section{Conclusions}
\label{sec:conc}

We put forth a novel production mechanism for magnetic monopoles arising from cosmic ray ISM interactions, leading to an irreducible astrophysical flux of monopoles. Unlike cosmological monopole production scenarios, this monopole flux is independent of primordial monopole abundances. This offers a persistent source of monopoles with variety of consequences for observations.
The presence of such monopoles could significantly hinder the formation and sustainability of galactic magnetic fields. 

We have generalized the conventional Parker bound—based on monopoles not disrupting present galactic magnetic fields—and the extended Parker bound—concerning monopoles not impeding the dynamo amplification of initial seed magnetic fields - to incorporate relativistic monopoles and monopole fluxes with extended energy spectra, providing a more comprehensive framework. By applying such generalized Parker-like and extended Parker-like bounds to monopole fluxes produced via cosmic ray ISM interactions, we derived new constraints on magnetic monopoles, independent of cosmology, based on the magnetic fields in the Milky Way and Andromeda galaxies. Our limits are competitive with other experimental techniques for monopoles with masses around 
$\sim 100$~GeV.
These results represent the first constraints on persistent astrophysical monopole fluxes derived from the destruction of galactic magnetic fields that do not depend on primordial, cosmogenic monopoles. Additionally, we highlight how cosmic ray ISM interactions open new avenues to probe magnetic dipoles and their effects on galactic magnetic fields.

Our scenario also provides an intriguing avenue to probe monopoles with magnetic charge less than unity, particularly in the parameter space below $\sim 100$~GeV, where some of the existing limits could be alleviated. Furthermore, we predict a persistent flux of magnetic monopoles that could be targeted by direct detection experiments, offering a compelling testing ground for novel monopole signatures. This motivates further investigations of our novel scenario.

\vspace{1cm}
{\bf Acknowledgments.}
The work of S.I. is supported by the JSPS KAKENHI Grant No.\,24K22879 and JPJSCCA20200002.
IS and VM are supported by the U.S. Department of Energy under the award number DE-SC0020250 and DE-SC0020262. 
V.T. acknowledges support from JSPS KAKENHI grant No. 23K13109 and by the World Premier International Research Center Initiative (WPI), MEXT, Japan.

\newpage 

\appendix
\section{Relativistic Monopole Energy Gain}
\label{app}

We derive here Eq.~\eqref{eq:monoE} is for energy gain of magnetic monopoles considering relativistic effects. We consider produced monopole flux from cosmic ray ISM interactions to be approximately isotropic and resulting in equal amount of monopoles and anti-monopoles. 

We can expect that the monopole distribution average $\langle dT/dt \rangle = 0$, where $T = (1/2) m \beta^2$ is the monopole kinetic energy with velocity $\vec{\beta}$.
Note that this assumption is conservative and the resulting bound is stronger without it.
Hence, the final kinetic energy of monopoles $T_f = T (t = \Delta t)$ after they traversed a magnetic field region characterized by $l_c$ over time $\Delta t \simeq l_c/\beta$ is
\be
\langle T_{f} \rangle  = \langle T_{i} \rangle + \frac{1}{2}(\Delta t)^{2} \frac{d^{2}T}{dt^{2}}~,
\ee
where $T_i = T(t = 0)$ is the initial monopole energy.

Consider non-relativistic monopole motion in a constant magnetic field $\vec{B}$ 
\be \frac{d \vec{p}}{dt} = m \frac{d\vec{\beta}}{dt} = g\vec{B}~.
\ee
Then, we find that 
\be 
\dfrac{dT}{dt} = m \vec{\beta} \cdot \dfrac{d\vec{\beta}}{dt} = g\vec{B} \cdot \vec{\beta},
\ee
where we have used the equation of motion. Differentiating again, we have
\be
\frac{d^{2}T}{dt^{2}} = \frac{d}{dt}(g\vec{B} \cdot \vec{\beta}) = g \left( \vec{\beta} \cdot \frac{d \vec{B}}{dt} + \vec{B}\cdot \frac{d \vec{\beta}}{dt} \right)~.
\ee
Since the magnetic field is assumed to be constant, we obtain 
\be
\frac{d^{2}T}{dt^{2}} =  g\vec{B}\cdot \frac{d \vec{\beta}}{dt} = \frac{g^{2} B^{2}}{m}~.
\ee
Therefore, the average energy gained by a monopole is
\be 
\Delta E \simeq \frac{1}{2} (\Delta t)^{2} \frac{d^{2}T}{dt^{2}} = \frac{1}{2}  \frac{g^{2} B^{2} l_c^2}{m \beta^2}~.
\label{eq:nonreldelE}
\ee

Let us now consider relativistic monopoles. First notice that for the magnetic field pointing in the $z-$direction the equations of motion can be written as
\be
\frac{dp_{z}}{dt} = gB, ~~~\frac{dp_{\perp}}{dt} = 0~,
\ee
where $p_z$ is momentum along $z$ direction and $p_{\perp}$ denotes orthogonal direction momentum. From this, it follows that
\be 
p_{z}(t) = p_{z,0} + gBt, ~~~ p_{\perp} = {\rm const.}
\ee

Next we can examine the total energy of a monopole
\be E^{2} = m^{2} + |\vec{p}(t)|^{2} = m^{2} + p_{\perp}^{2} + p_{z}^{2}(t)~.
\ee
Hence
\be 
E^{2} = m^{2} + p_{\perp}^{2} + (p_{z,0} + gB t)^{2} = E_{0}^{2} + (2gBt)p_{z,0}+g^{2}B^{2} t^{2}~,
\ee
where $E_{0}^{2} \equiv m^{2} + p_{\perp}^{2} + p_{z,0}^{2}$ is the initial monopole energy.

Now we can find change in monopole energy after traversing a magnetic field region as $\Delta E \equiv E-E_{0}$ and write
\be 
E = E_{0} \sqrt{1+\left(\frac{2gBtp_{z,0}+g^{2}B^{2} t^{2}}{E_{0}^{2}}\right)}~.
\ee
Recalling that we are interested in the $E_{0} \gg gB l,~ E_{0} \gg gBt $ limit, we have
\be
E \simeq E_{0} \left(1 +\frac{1}{2} \left(\frac{2gBtp_{z,0}+g^{2}B^{2} t^{2}}{E_{0}^{2}}\right)\right) ~\Longrightarrow~         \Delta E = \frac{2gBtp_{z,0}+g^{2}B^{2} t^{2}}{2E_{0}}~.
\ee
Finally, as in analysis of Ref.~\cite{Turner:1982ag}, we can observe that the term linear in $g$ will average to zero in the presence of equal monopole and anti-monopole fluxes. As a result, considering $t = \Delta t \simeq l_c / \beta$, we obtain
\be 
\Delta E \simeq \frac{(gBt)^{2}}{2m\gamma} = \frac{(gBl_{c})^{2}}{2m \beta^{2} \gamma}, 
\ee
which reduces to Eq.~\eqref{eq:nonreldelE} in the $\gamma \rightarrow 1$ limit.

\bibliography{ref}
\addcontentsline{toc}{section}{Bibliography}
\bibliographystyle{JHEP}
\end{document}